\documentclass[10pt,conference]{IEEEtran} 

\newcommand{\scheme}{VisualRepair}

\usepackage{enumitem}
\usepackage{multirow}
\usepackage[table]{xcolor}
\usepackage{tcolorbox}
\usepackage{url}
\usepackage{algorithmic}  
\usepackage[ruled, linesnumbered]{algorithm2e}

\IEEEoverridecommandlockouts

\usepackage{cite}
\usepackage{amsmath,amssymb,amsfonts}
\usepackage{algorithmic}
\usepackage{graphicx}
\usepackage{textcomp}
\usepackage{xcolor}
\usepackage{booktabs}
\def\BibTeX{{\rm B\kern-.05em{\sc i\kern-.025em b}\kern-.08em
    T\kern-.1667em\lower.7ex\hbox{E}\kern-.125emX}}
\begin{document}


\title{\raisebox{-0.5ex}{\includegraphics[width=1.1cm]{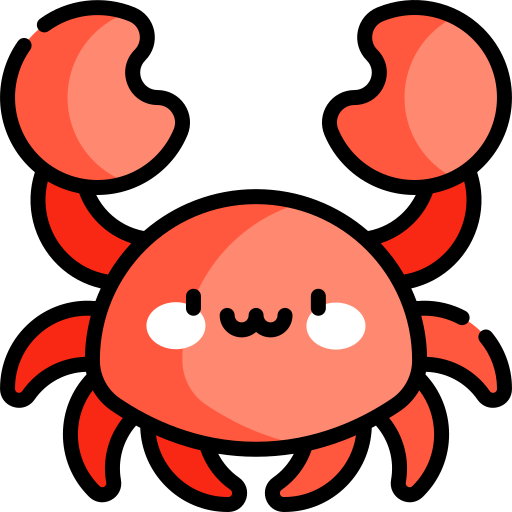}} \scheme: Dynamic Tool Calling and Region Focusing for Visual Software Issue Repair}

\author{\IEEEauthorblockN{Anonymous Author(s)}
}

\author{%
  Jingyu Xiao$^\dagger$, 
  Zhongyi Zhang$^{\dagger}$, 
  Haoran Hou$^\dagger$,
  Yuxuan Wan$^\dagger$,
  Yuan Jiang$^\P$,
  \\
  Yintong Huo$^{\ddagger\mathsection}$\thanks{$^{\mathsection}$Corresponding authors: Yintong Huo (ythuo@smu.edu.sg)},
  Michael R. Lyu$^\dagger$ \\
  $^\dagger$The Chinese University of Hong Kong, Hong Kong SAR, China \\ 
  $^\P$Harbin Institute of Technology, China\\
  $^\ddagger$Singapore Management University, Singapore\\
  \IEEEauthorblockA{ 
  \{jyxiao, zyzhang, hrhou, yxwan\}@link.cuhk.edu.hk, jiangyuan@hit.edu.cn, \\ ythuo@smu.edu.sg, lyu@cse.cuhk.edu.hk}
}


\maketitle

\begin{abstract}
Automated Program Repair (APR) has witnessed significant progress with the advent of Large Language Models (LLMs). However, as modern software systems increasingly expose rich graphical user interfaces, effectively leveraging visual information from bug screenshots has become essential for understanding bugs and generating accurate fixes in multimodal scenarios. Real-world issue reports frequently contain heterogeneous visual attachments including UI screenshots, IDE snapshots, GIFs, and text-centric images, each with distinct visual patterns and domain-specific semantics that impose substantial perceptual demands on MLLMs. Furthermore, bug screenshots often contain large expanses of uninformative and bug-irrelevant regions, distracting the model's attention and limiting patch diversity.  To address these challenges, we propose \scheme, an MLLM-based framework for visual software issue repair comprising two core modules: Image Type-aware Tool Calling (ITTC), which classifies input images and dynamically invokes a tailored tool-calling chain for robust visual interpretation, and Dynamic Test-time Region Focusing (DTRF), which grounds multiple bug-related region candidates and refines them via an adaptive zoom-in and zoom-out strategy to improve fault localization and promote diverse patch generation. Extensive experiments on the SWE-bench Multimodal benchmark demonstrate that \scheme\ consistently outperforms state-of-the-art approaches. \scheme\ resolves 196 and 25 instances on the test and dev sets, respectively, surpassing the best baseline by 10 and 11 instances. These results highlight the effectiveness of type-aware visual understanding and region-focused localization for automated visual software issue repair.

\end{abstract}

\footnotetext[1]{The paper was completed in March 2026 and is currently under review. The code will be released upon acceptance.}


\begin{IEEEkeywords}
Automated Program Repair, MLLMs, GUI.
\end{IEEEkeywords}

\section{Introduction}

Software bugs are inevitable in modern software development, posing significant threats to system reliability, security, and user experience. Automated Program Repair (APR)~\cite{zhang2023survey, gazzola2018automatic} has emerged as a promising research direction aimed at automatically localizing and fixing software defects, thereby reducing the manual effort required for software maintenance and improving overall software quality. Traditional APR techniques~\cite{le2011genprog, liu2019tbar, nguyen2013semfix} have achieved notable progress, yet they often struggle with generalizability and the complexity of real-world bugs. With the advent of LLMs, APR has witnessed significant progress via fine-tuning~\cite{LLM4APR_Huang_ASE, LLM4APR_Jiang, FitRepair, MoRepair} and prompting-based paradigms~\cite{AlphaRepair, GAMMA, xia2024chatrepair, ThinkRepair, RepairAgent}, achieving promising results on benchmarks like Defects4J~\cite{just2014defects4j} and SWE-bench~\cite{jimenez2024swebench}. These approaches typically follow a generate-and-validate paradigm, where candidate patches are generated from bug reports and subsequently verified through test execution~\cite{xia2025demystifying, yang2024sweagent, SpecRover}.

However, most modern software systems, including web applications, design tools, and integrated development environments (IDEs), expose rich graphical user interfaces (GUIs) to users for display and interaction. Bugs in these systems frequently manifest as visual anomalies~\cite{xiao2025designbench}, layout defects, or interaction failures~\cite{xiao2025interaction2code} that are inherently difficult to describe through text alone, and thus cannot be adequately captured by existing text-only methods. To bridge this gap, SWE-bench Multimodal (SWE-bench M)~\cite{yang2025swebench} is proposed, extending the SWE-bench framework to incorporate visual contexts such as screenshots alongside natural language issue descriptions. Because of advanced visual understanding and code generation capabilities~\cite{zhao2026beyond, tang2025slidecoder,tang2026efficientpostergen, teoh2026webtestpilot,le2026uibenchkit}, Multimodal Large Language Models (MLLMs) have been applied to visual software issue repair by perceiving GUI-level defects~\cite{GUIPilot} directly from screenshots, reasoning about the underlying root causes, and generating effective patches. For example, DesignRepair~\cite{yuan2025designrepair} proposes a design-guideline-aware system that analyzes and repairs design issues for UIs. GUIRepair~\cite{huang2025seeing} designs a bidirectional pipeline including image2code and code2image to achieve cross-modality reasoning for visual software issue repair.




Despite the promise of MLLMs for visual software issue repair, there remain two challenges  hindering effectiveness.

\begin{figure}[t]
    \centering
    \includegraphics[width = .49\textwidth]{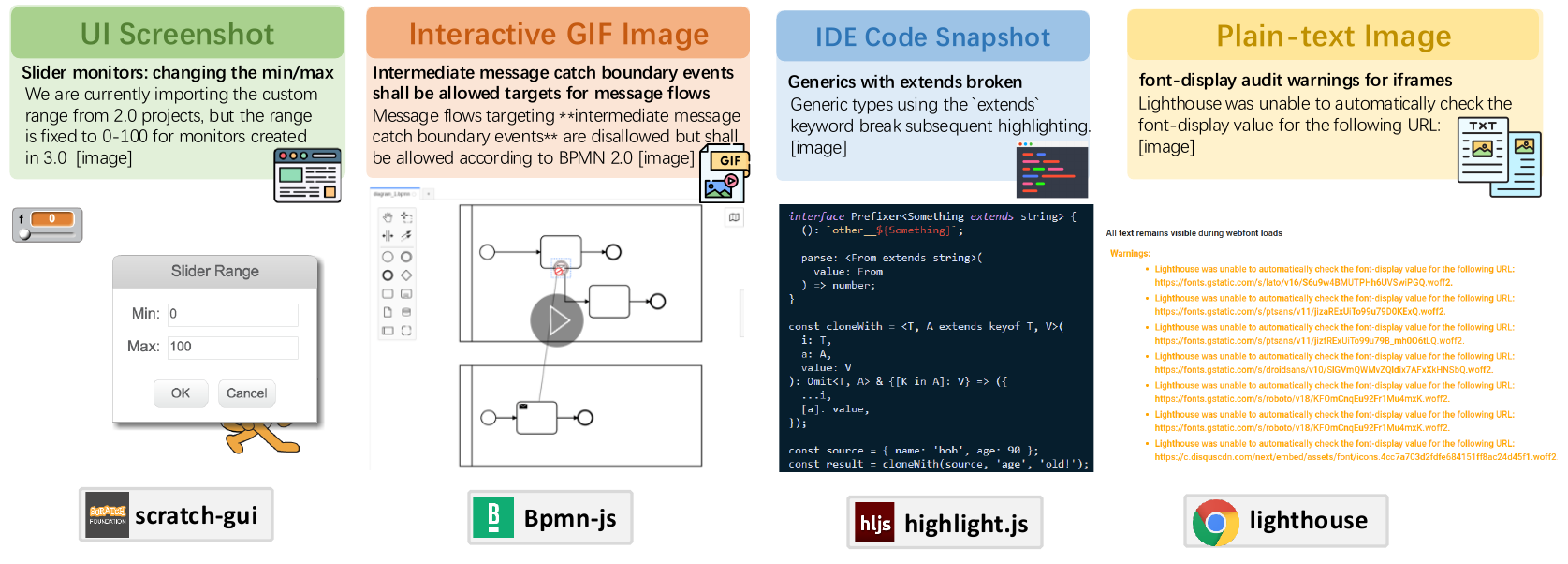}
    \caption{Four visual tasks with heterogeneous inputs for SWE-bench M~\cite{yang2025swebench}.}
    \label{fig:example}
\end{figure}

\textbf{Challenge 1: Heterogeneous visual content in issue reports hinders MLLMs’ ability to understand bugs.} Visual artifacts in real-world software issue reports are heterogeneous, spanning UI screenshots, animated GIFs, Integrated Development Environment (IDE) code snapshots, and plain-text images as shown in Figure~\ref{fig:example}. This diversity imposes varied aspects on image reasoning, as each modality exhibits distinct visual patterns and domain-specific semantics. For instance, GIFs require temporal reasoning over interaction sequences, while plain-text images demand accurate text extraction and interpretation. Notably, UI-based and IDE-based inputs pose different challenges for bug reproduction: UI inputs are highly variable and require inferring interaction logic, whereas IDE inputs are more structured but demand a precise understanding of code semantics. Treating these heterogeneous inputs in a uniform manner will significantly hamper MLLMs’ ability to comprehend and resolve reported issues.




\textbf{Challenge 2: MLLMs fail to locate bug-related regions when facing complex bug screenshots.} Real-world bug screenshots encompass overwhelming yet irrelevant information: they usually contain a large portion of uninformative content (e.g., blank backgrounds) for debugging. Such visual noise poses two compounding problems: 1) it distracts the model's attention from the defective region, leading to imprecise fault localization; 2) processing full, unfiltered images constrains the model’s ability to examine the bug from multiple perspectives, thereby limiting the diversity of generated patches.

In this paper, we propose \scheme, an MLLM-based framework for visual software issue repair that addresses the aforementioned challenges. \scheme \ comprises two core modules: Image Type-aware Tool Calling (ITTC) and Dynamic Test-time Region Focusing (DTRF).
To handle the heterogeneity of visual inputs in real-world issue reports, ITTC employs a lightweight classifier to automatically identify the type of each input image and dynamically assign a tailored tool-calling chain accordingly. The toolkit consists of four components: a GIF tool that processes animated images and extracts key frames to capture dynamic bug behaviors; an OCR tool that extracts textual content from image regions; a cropping tool that removes uninformative blank areas to produce a more compact visual representation; and a code library tool containing code templates that enriches the context for bug reproduction and patch selection.
To address the challenge of imprecise bug localization, we further propose DTRF, which mitigates noisy visual context and promotes diverse patch generation. DTRF first prompts the MLLM to identify and ground multiple bug-related region candidates within the input image. For each candidate region, a zoom-in and zoom-out strategy is then applied to dynamically adjust the region boundary, enabling the model to refocus on the most diagnostically relevant area at an appropriate level of granularity. 
Compared with a range of state-of-the-art methods in the SWE-bench M leaderboard~\cite{Leaderboard}. \scheme\ resolves \textbf{196} and \textbf{25} issues on the test and dev splits, respectively, surpassing the best baselines by \textbf{10} and \textbf{11} instances on test and dev splits, respectively, as of March 2026. Our contributions are as follows:

\begin{itemize}[leftmargin=*]

    \item  \textbf{Insights}. We identify two critical yet underexplored phenomena in visual software issue repair: the heterogeneity of visual inputs in real-world issue reports and imprecision of bug-related visual region localization.

    \item  \textbf{Technique.} We propose \scheme, an MLLM-based framework that performs type-aware tool invocation for heterogeneous visual understanding and multi-region grounding with zoom-in and out refinement for precise fault localization and diverse patch generation.

    \item \textbf{Evaluation.} We conduct extensive experiments on the SWE-bench M benchmark~\cite{yang2025swebench}, comparing \scheme\ against a range of state-of-the-art methods in the leaderboard~\cite{Leaderboard}. \scheme\ resolves 196 and 25 issues on the test and dev sets, respectively, outperforming all existing baselines.



\end{itemize}



\section{Background}

\subsection{Problem Definition}

The visual software issue repair task is formulated as follows. The input consists of two components: a software repository $\mathcal{R} = \{\mathcal{C}, \mathcal{D}\}$, comprising the project codebase $\mathcal{C}$ and associated documentation $\mathcal{D}$, and an issue report $\mathcal{I} = \{\mathcal{T}, \mathcal{V}\}$, consisting of a natural language bug description $\mathcal{T}$ and a set of visual attachments $\mathcal{V} = \{v_1, v_2, \dots, v_k\}$. Each image $v_i \in \mathcal{V}$ is associated with a type $\tau_i \in \{\texttt{UI}, \texttt{IDE}, \texttt{GIF}, \texttt{Text}, \dots\}$, reflecting the heterogeneous nature of visual content found in real-world issue reports. The objective is to generate a patch $\mathcal{P} = \{(f_i, \Delta_i)\}$, where each pair denotes a code modification $\Delta_i$ applied to source file $f_i \in \mathcal{C}$, such that $\mathcal{P}$ resolves the reported defect without introducing regressions. The task is formulated as $\mathcal{M}: (\mathcal{R}, \mathcal{I}) \rightarrow \mathcal{P}$, where $\mathcal{M}$ is the multimodal large language model.


\subsection{Preliminary Study}

\subsubsection{The Image Type Distributions} 
\begin{figure}[ht]
    \centering
    \includegraphics[width = .45\textwidth]{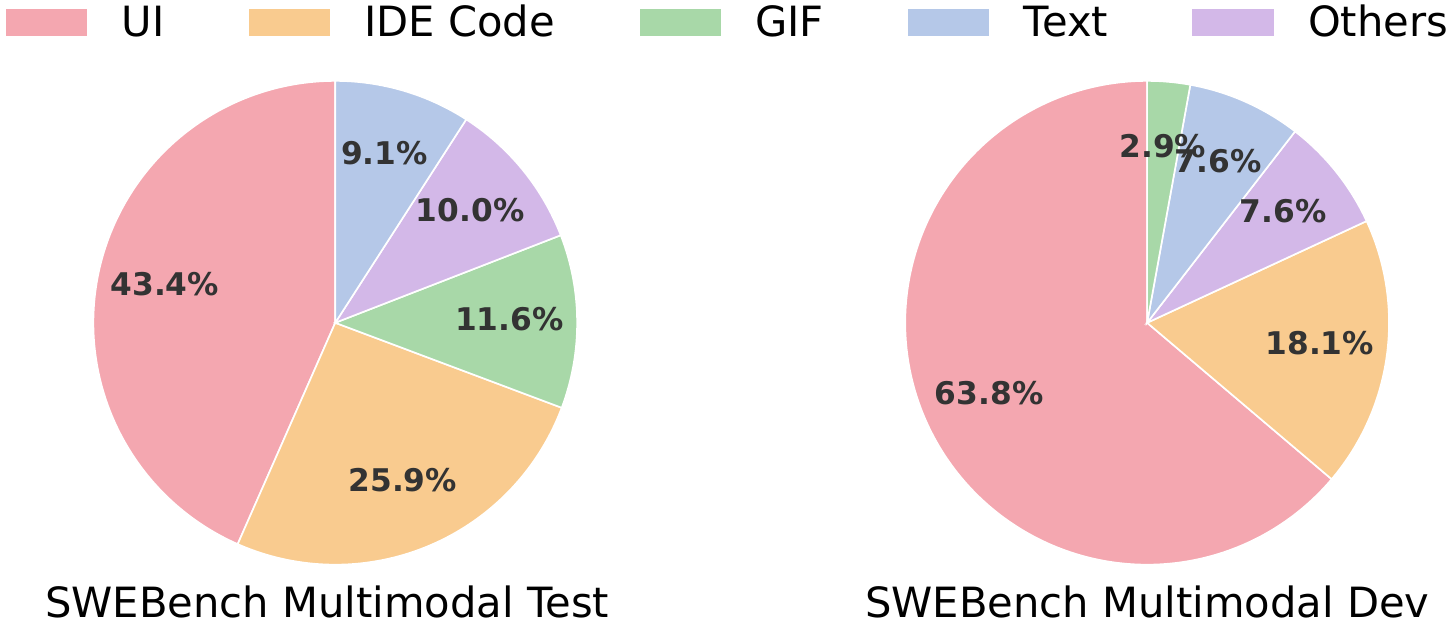}
    \caption{Image types distribution in SWEbench MM benchmark. Others means no image or bug-unrelated image only.}
    \label{fig:image_type_distribution}
\end{figure}

We employ four PhD students majoring in software engineering with comprehensive front-end development background to annotate the bug images in SWE-bench Multimodal benchmark~\cite{yang2025swebench}. We first randomly select 10\% issues for analysis and then discuss and refine the image type until reaching a consensus. The annotation results are presented in Fig.~\ref{fig:image_type_distribution}. UI screenshots constitute the dominant image type, accounting for 43.4\% and 63.8\% of issues in the test and dev splits respectively, followed by IDE code snapshots at 25.9\% and 18.1\%, GIF images at 11.6\% and 2.9\%, and plain text images at 9.1\% and 7.6\%. The remaining 10.0\% and 7.6\% fall into the others category, comprising issues with no attached image or bug-irrelevant visual content.


\begin{tcolorbox}[colback=gray!20, colframe=gray!20, width=\columnwidth, left=0.05in, right=0.05in, top=0.05in, bottom=0.05in]
\textbf{Observation 1:} Visual inputs in real-world issue reports are highly heterogeneous, spanning four distinct types, underscoring the necessity of diverse perception strategies.
\end{tcolorbox}


\subsubsection{Inaccurate Bug-related Region Focusing} 
\begin{figure}[ht]
    \centering
    \includegraphics[width = .42\textwidth]{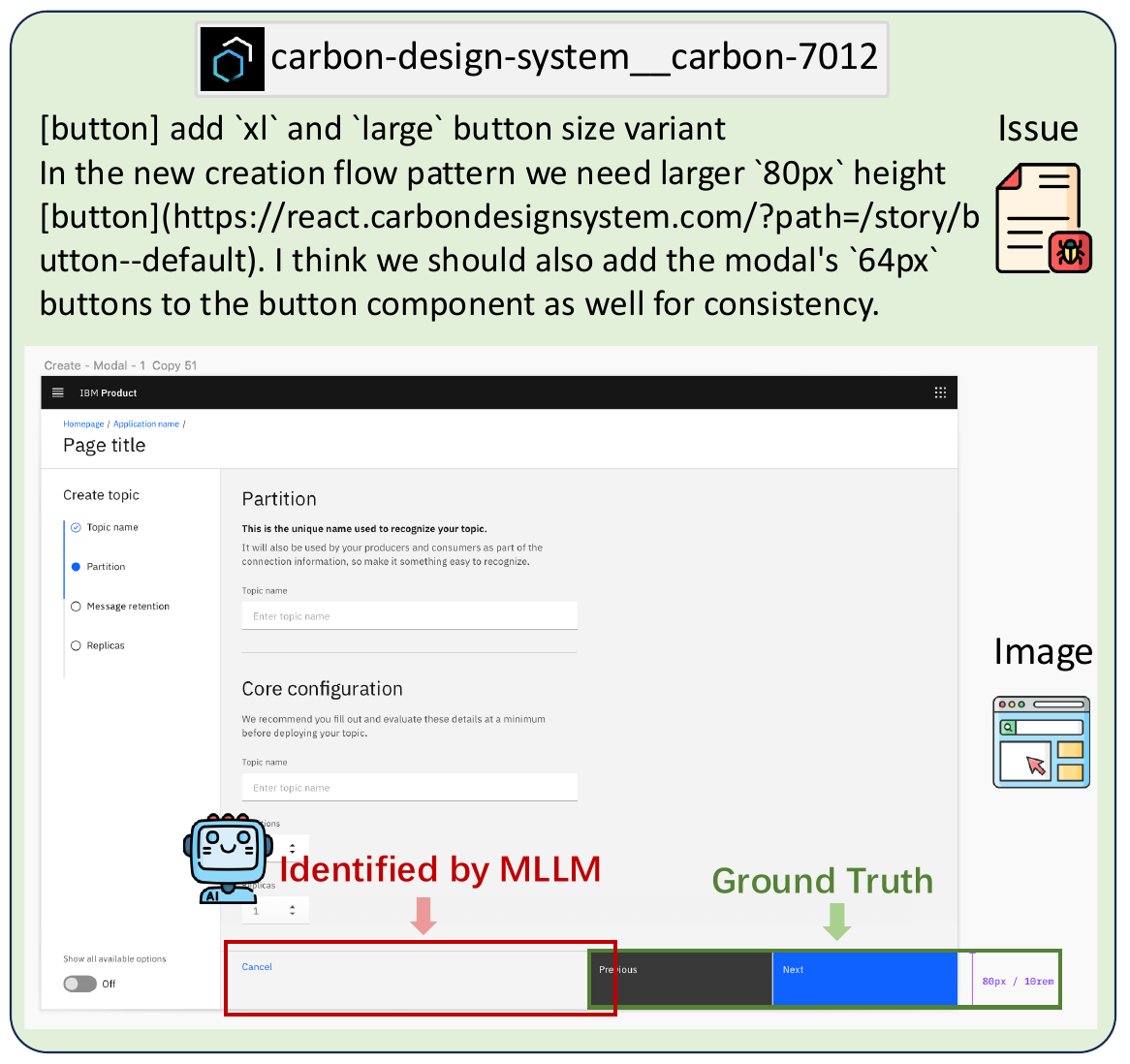}
    \caption{Bug-related region grounding example.}
    \label{fig:ground_example}
\end{figure}

Even when an MLLM successfully processes the input image, precisely localizing the bug-relevant region remains a significant challenge. As illustrated in Figure~\ref{fig:ground_example}, for issue \texttt{carbon-7012}~\cite{carbon-design-system}, the reported bug requests the addition of \texttt{xl} and \texttt{large} button size variants with an \texttt{80px} height. While the ground truth region (green box) correctly identifies the full button row at the bottom of the UI, the MLLM-identified region (red box) only partially overlaps with it, missing the rightmost portion containing the key size indicator (\texttt{80px / 10rem}). This demonstrates that MLLMs have limitations in capturing the full extent of the bug-relevant region, particularly when the defective area is located at the periphery of a complex, content-rich screenshot. 


\begin{tcolorbox}[colback=gray!20, colframe=gray!20, width=\columnwidth, left=0.05in, right=0.05in, top=0.05in, bottom=0.05in]
\textbf{Observation 2:} MLLMs exhibit imprecise grounding capability, frequently failing to localize the bug-relevant region, which undermines  repairing performance.
\end{tcolorbox}

\begin{figure*}[ht]
    \centering
    \includegraphics[width = .98\textwidth]{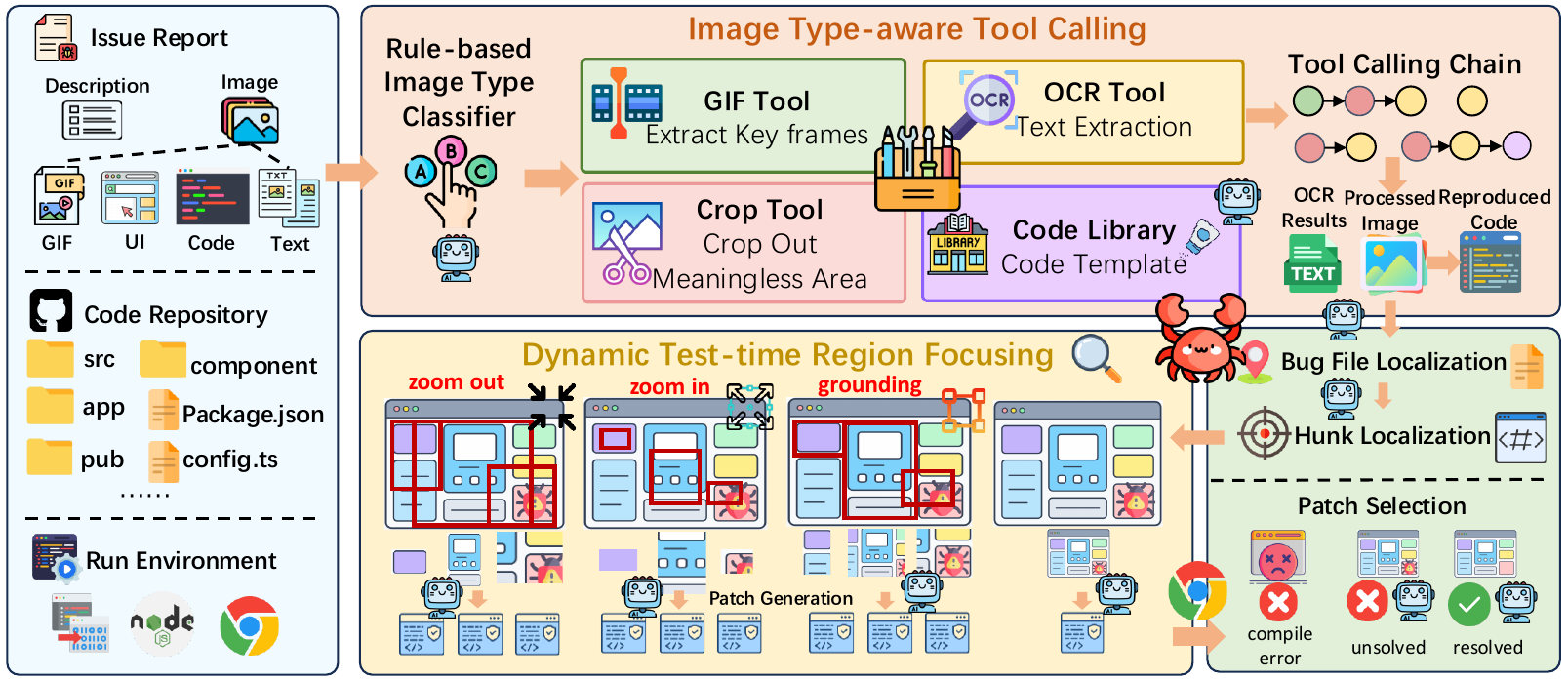}
    \caption{The overview of \scheme.}
    \label{fig:overview}
\end{figure*}

\section{Method}

\subsection{Overview}

Fig.~\ref{fig:overview} illustrates the overall architecture of \scheme. Given an issue report $\mathcal{I}$ and a software repository $\mathcal{R}$, \scheme \ proceeds through three stages. First, the Image Type-aware Tool Calling module classifies each input image and routes it through a tailored tool-calling chain, producing structured outputs including OCR results, a processed image, and reproduced code. Second, \scheme \ performs hierarchical fault localization on $\mathcal{R}$, identifying the bug-relevant files and code hunks to narrow the repair search space. Finally, the Dynamic Test-time Region Focusing module grounds some bug-related region candidates from the processed image, augments each via zoom-in and zoom-out operations, and generates candidate patches. Finally, patch selection is conducted based on the compilation validation and visual difference checking.


\subsection{Image Type-aware Tool Calling}

\subsubsection{Tool sets}

Applying a uniform perception strategy across visually heterogeneous inputs causes MLLMs to misinterpret GIF interaction sequences, overlook code-level details in IDE snapshots, and fail to extract critical text from text-centric images, ultimately hindering bug comprehension. Therefore, we implement the following tools:


\textbf{Gif tool}. Real-world issue reports include animated GIFs to demonstrate dynamic or interactive bug behaviors that cannot be captured by a single static screenshot. Directly feeding a GIF to an MLLM is infeasible; instead, we decompose each GIF into a sequence of individual frames, producing an ordered image list. However, retaining all frames is redundant, as consecutive frames in a GIF are often visually near-identical. We extract keyframes from a GIF by measuring the mean absolute error (MAE) between consecutive preprocessed frames. A dynamic threshold is defined as $\tau = \mu_{\text{MAE}} + k \cdot \sigma_{\text{MAE}}$, where $\mu_{\text{MAE}}$ and $\sigma_{\text{MAE}}$ are the mean and standard deviation of all inter-frame MAEs, and $k$ is a scaling coefficient. Starting from the first frame, a subsequent frame is selected as a keyframe if its MAE relative to the last selected keyframe exceeds $\tau$, subject to a minimum frame gap constraint to avoid redundant selections.



\textbf{OCR tool}. Some issue images, particularly IDE snapshots and terminal outputs, contain dense bug-relevant text such as error messages, stack traces, or code excerpts. We employ PaddleOCR~\cite{cui2025paddleocr30technicalreport} to detect and transcribe all text regions within the input image, and incorporate the extracted text into the MLLM's input context as a complementary signal, compensating for cases where direct visual interpretation of small or low-contrast text may be unreliable.

\textbf{Crop tool}. Bug screenshots often contain large uninformative regions like blank backgrounds or empty canvas areas that contribute scarce information yet inflate the image token count and distract the model's attention. To address this, we apply a structure-aware cropping procedure based on the UIED~\cite{xie2020uied} algorithm, which detects UI elements within the image, representing each detected element as a bounding box $b_i$. We then perform a horizontal and vertical scan-line sweep across the image: a scan line that does not intersect any bounding box identifies a region devoid of UI elements. Such regions are marked as removable and subsequently cropped from the image. The resulting image retains only the regions containing meaningful UI content, yielding a more compact representation that reduces unnecessary token consumption and sharpens the model's focus on bug-relevant areas.

\begin{figure}[t]
    \centering
    \includegraphics[width = .49\textwidth]{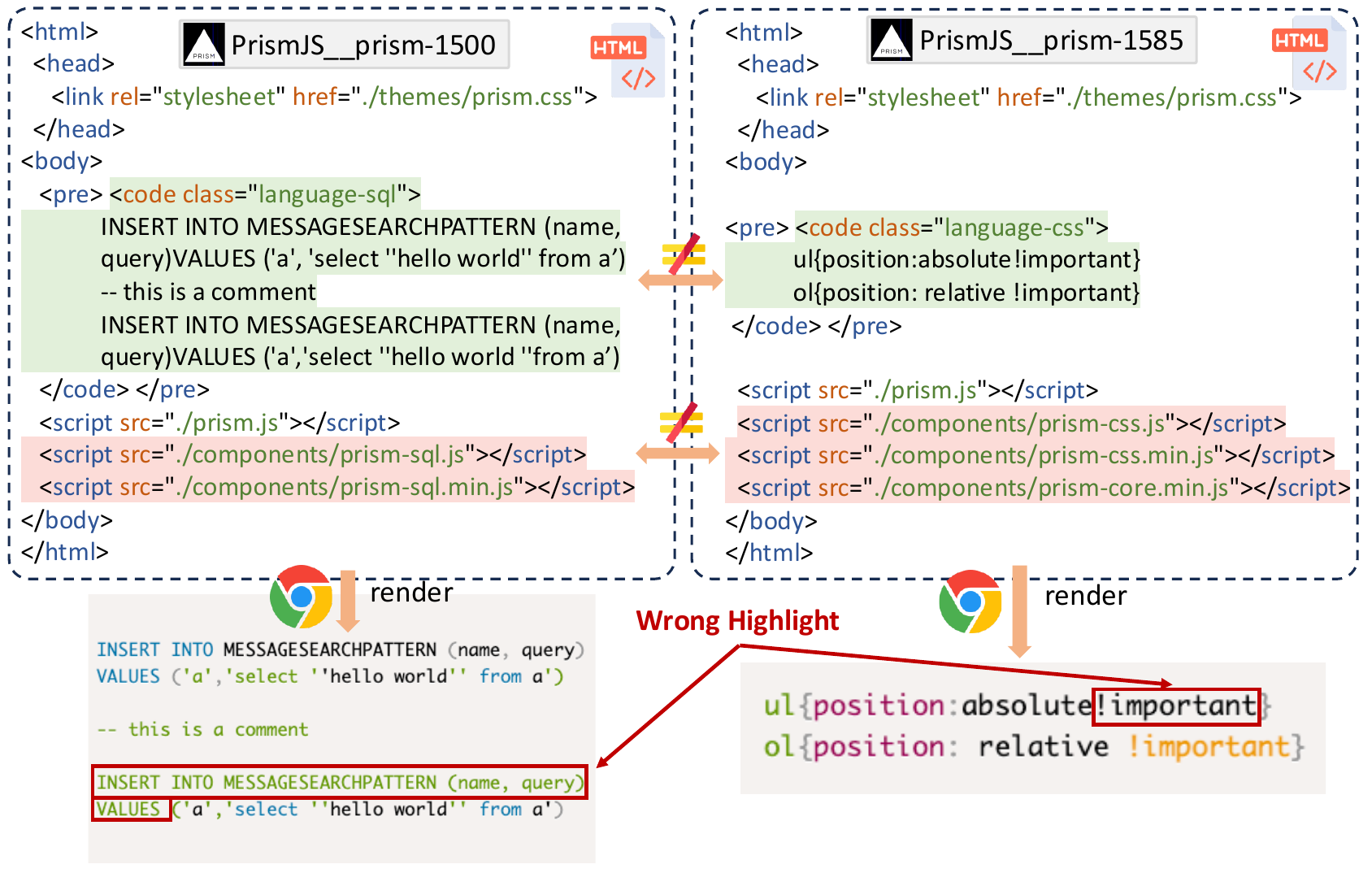}
    \caption{The reproduced code of two issues of PrismJS.}
    \label{fig:template_example}
\end{figure}

\textbf{Code library}. Front-end libraries concerned with format rendering, such as syntax highlighting libraries (e.g., PrismJS~\cite{PrismJS}, Highlight.js~\cite{Highlight}) and Markdown rendering libraries (e.g., Marked.js~\cite{Marked}), exhibit a high degree of structural homogeneity in their code. As illustrated in Figure~\ref{fig:template_example}, bugs in such libraries typically manifest as incorrect visual rendering behavior (e.g., wrong syntax highlighting regions), yet the underlying HTML code required to reproduce them follows a near-identical template: a standard HTML scaffold that imports the relevant JavaScript dependencies, and embeds a minimal code snippet to trigger the erroneous behavior. Reproducing a bug in these libraries reduces to filling in two components within a shared template: the triggering code content and the corresponding library import statements.

Motivated by this observation, we construct a Code Library of code templates. Templates are constructed on-demand and issue-agnostically: when a previously unseen repository is encountered, an MLLM analyzes its structure and official documentation to generate a minimal runnable template. Since construction relies solely on repository-level information with no access to issue content, this process eliminates any risk of test-set leakage. When the MLLM is tasked with reproducing a bug, the Code Library Tool retrieves the matching template and populates it with issue-specific content, significantly reducing the burden on the model to generate correct reproduction code from scratch. This not only improves the fidelity of bug reproduction, but also enables automated patch validation: once a candidate patch is applied, the resulting HTML file is rendered by a WebDriver, which captures a screenshot and compares it against the original issue image to assess whether the visual defect has been resolved.

\subsubsection{Image Type Classifier}

\begin{algorithm}
\caption{Image Type-aware Tool Calling}
\label{alg:ittc}
\begin{algorithmic}[1]
\REQUIRE Input image $v$, file suffix $s$, issue $\mathcal{I}$
\ENSURE Tool calling result $\mathcal{O}$

\STATE $repo \leftarrow \textsc{GetRepo}(\mathcal{I})$
\IF{$repo \notin \textsc{CodeLibrary}$}
    \STATE $\text{code\_template} \leftarrow \textsc{GenerateTemplate}(repo)$
    \STATE $\textsc{CodeLibrary} \leftarrow \textsc{CodeLibrary} \cup \{repo: \text{code\_template}\}$
\ENDIF

\IF{$s = \texttt{.gif}$}
    \STATE $\text{frames} \leftarrow \textsc{GifTool}(v)$
    \STATE $\text{key\_frames} \leftarrow \textsc{KeyFrameExtraction}(\text{frames}, \delta)$
    \STATE $code \leftarrow MLLM(\text{key\_frames}, \text{bug\_description})$
    \STATE $\mathcal{O} \leftarrow (code, \text{key\_frames}, \text{bug\_description})$
\ELSE
    \STATE $\mathcal{E} \leftarrow \textsc{UIED}(v)$
    \COMMENT{Detect UI elements with bounding boxes}
    \STATE $v' \leftarrow \textsc{CropTool}(v, \mathcal{E})$
    \COMMENT{Remove uninformative regions}
    \IF{$\exists\ e \in \mathcal{E}\ \text{s.t.}\ \textsc{Type}(e) \neq \texttt{text}$}
        \STATE $code \leftarrow MLLM(v', \text{bug\_description})$
    \ELSE
        \STATE $\text{text\_content} \leftarrow \textsc{OCRTool}(v')$
        \COMMENT{Extract textual content}
        \IF{$\textsc{IsCode}(\text{text\_content})$}
            \STATE $\text{code\_template} \leftarrow \textsc{CodeLibrary}[repo]$
            \STATE $code \leftarrow MLLM(\text{code\_template}, v', \text{bug\_description})$
            \STATE $\mathcal{O} \leftarrow (code, v', \text{bug\_description})$
        \ELSE
            \STATE $\mathcal{O} \leftarrow (\text{text\_content}, \text{bug\_description})$
        \ENDIF
    \ENDIF
\ENDIF
\RETURN $\mathcal{O}$
\end{algorithmic}
\end{algorithm}

Algorithm~\ref{alg:ittc} summarizes the Image Type-aware Tool Calling procedure, which routes each input image through a sequential chain based on its associated types. The classification begins with a lightweight check on the file suffix $s$. If the image is a GIF, it is passed to the GIF Tool, which decomposes the animation into a frame sequence and applies MAE-based key frame extraction to obtain a compact set of representative frames. The MLLM then takes the key frames together with the bug description as input to generate the reproduction code, and the final output $\mathcal{O}$ packages the generated code, key frames, and bug description for downstream processing.


For non-GIF images, UIED~\cite{xie2020uied} detects all UI elements as bounding boxes, and the crop tool is immediately applied to remove blank regions, yielding a compact representation $v'$. If any non-text UI element is detected, the image is treated as a UI screenshot and the MLLM generates reproduction code from $v'$ and the bug description. Otherwise, OCR extracts text from $v'$, which is then classified (\textsc{IsCode}) as code or plain text. For code images, a matching template is retrieved and combined with $v'$ and the bug description to prompt code generation. For plain-text images, the extracted text and bug description are directly returned as bug-related ingredients $\mathcal{O}$ for subsequent localization and repair.

\textbf{Performance of Image Classifier}. The classification in Algorithm~\ref{alg:ittc} is not a learned model but a deterministic rule-based pipeline, which inherently avoids the misclassification risks of trained classifiers. It proceeds in three stages: (1) GIF detection is performed purely by file suffix, yielding no errors by definition; (2) UI element routing is handled by UIED's rule-based bounding box detector, where any detected non-text UI element triggers the UI branch; (3) only the final text-vs-code discrimination relies on an MLLM judgment over the extracted text content. Since stages (1) and (2) cover the majority of inputs deterministically, the MLLM (e.g., GPT-4o) is only invoked for a small, well-defined subset. We empirically evaluated the end-to-end classification accuracy on the full SWE-bench Multimodal benchmark and confirmed 100\% accuracy with zero misclassified instances, demonstrating that the cascading error risk does not manifest in practice.

The differences between ITTC and general tool-calling frameworks. Unlike ReAct~\cite{yao2023react}, OpenHands~\cite{wang2024openhands}, and SWE-agent~\cite{yang2024sweagent}, which rely on free-form LLM reasoning for tool selection, ITTC adopts a deterministic, type-driven routing mechanism that eliminates the instability and overhead of LLM-based tool dispatch for visual inputs. 


\subsection{Bug Localization}

Following Agentless~\cite{xia2025demystifying} and GUIRepair~\cite{huang2025seeing}, \scheme\ performs fault localization in two stages: file localization and hunk localization. In file localization, candidate files are identified by combining MLLMs and embedding models. The MLLM examines the repository structure given project documentation and the enhanced issue report to return likely buggy file paths, avoiding full file contents to stay within context limits. Complementarily, the embedding model retrieves the Top-$N$ most semantically relevant files within MLLM-identified directories. The two candidate sets are merged, and if the union exceeds $N$, each file is compressed into a lightweight skeleton retaining imports, signatures, and comments, from which the MLLM selects the $N$ most relevant bug files. In hunk localization, the full contents of all key bug files are provided to the MLLM, which identifies suspicious classes, functions, or global constructs. Localization operates at class or function granularity to preserve contextual information. Elements exceeding $N$ lines are compressed to their headers, variable declarations, and comments. The MLLM returns suspected element names, and their complete bodies are extracted as the localized hunks.


\subsection{Dynamic Test-time Region Focusing}

Precisely identifying the bug-relevant region within an issue image is critical to generate accurate patches, yet a single localization attempt is often insufficient due to the limitation of MLLMs' visual grounding abilities. Therefore, we propose Dynamic Test-time Region Focusing, which augments patch generation with diverse, multi-granularity region candidates derived from the issue image.

\subsubsection{Bug Region Grounding} Given an issue report consisting of issue descriptions and issue images, we employ a grounding MLLM to localize the bug-relevant region within the input image $v$, producing bounding boxes that delineate the suspected defective area. Since a single grounding result may be imprecise due to visual complexity or model uncertainty, we prompt the grounding agent to generate $M$ candidate regions $\{b_1, b_2, \dots, b_M\}$ for each image, introducing more diversity.



\subsubsection{Zoom-in and Zoom-out} To capture bug-relevant context at varying granularities, we apply zoom-in and zoom-out strategies to each of the $M$ grounded regions. For each candidate bounding box $b_i$, we produce two augmented variants: a zoomed-in region cropped to half the original area for fine-grained details, and a zoomed-out region expanded to twice the original area for broader context. This yields $2M$ additional candidates, resulting in $3M + 1$ distinct visual inputs.

\subsubsection{Patch Generation} Each of the $3M + 1$ region candidates, together with the class/function level bug code snippets identified by previous bug localization phase, is independently fed into the MLLM to generate patches. For each visual input, the MLLM produces $P$ candidate patches, yielding a final pool of $(3M + 1)\times P$ patches in total. This diverse patch pool, grounded in complementary spatial perspectives of the bug region, substantially increases the likelihood of generating a correct and complete repair. 
Specifically, each patch consists of a Search/Replace edit pair comprising two components: (i) a search segment containing the original code to be modified, and (ii) a replacement segment containing the revised code. Applying such a patch is straightforward: the search segment is located within the target file and substituted with the corresponding replacement. This edit-based patch format constrains the MLLMs to produce minimal, targeted modifications rather than complete rewrites and enhances reliability by lowering the risk of hallucinations that tend to arise when generating large code blocks from scratch.






\subsection{Patch Selection}




Given the pool of $(3M+1)\times P$ candidate patches generated by DTRF, we apply the following patch selection procedure to identify the most plausible repair. (1) Patch Filtering. We first attempt to apply and compile each candidate patch, recording whether it compiles successfully. Patches that fail to compile are excluded from subsequent stages, and only syntactically valid patches are retained. For issues associated with repositories that have a corresponding code template in our code library, we perform automated visual validation. Surviving patch are applied to the reproduction code, which is then rendered in a browser via web driver to capture a screenshot. Rendered screenshots are subsequently compared against the original issue images at the pixel level. Patches whose rendered output is visually identical to the buggy screenshot are discarded, as they indicate no effective change in visual behavior. (2) Patch Selection. We follow the GUIRepair~\cite{huang2025seeing} for patch selection, the remaining patch scenario images are then sequentially input to the MLLM, which assesses whether the visual output reflects a successful fix based on its understanding of the issue report. Once the model identifies a patch as effective, the validation process terminates. The validated patch is then retained as the resolution for the issue.



\section{Experiment Setup}

\subsection{Implementation}

 
\noindent \textbf{Implementation details.}  We adopt o3~\cite{o3} as the backbone MLLM for \scheme, and use \texttt{text-embedding-3-small}~\cite{text-embedding-3-small} as the embedding model for document retrieval. In the Image Type-aware Tool Calling module, the MAE threshold $\delta$ for GIF key frame extraction is set to 1.2, and the retrieval module returns the top-6 relevant documents from the repository for context augmentation. In the bug file localization phase, we set the sampling temperature to 1 with 2 sampling iterations, returning the top-4 most relevant bug files as candidates; in the hunk localization phase, we adopt the same temperature and sampling configuration to cover a diverse set of potential bug code snippets and provide adequate repair ingredients for patch generation. These retrieval and sampling hyperparameters are inherited directly from prior work (GUIRepair, Agentless) and kept consistent with their settings, which ensures fair comparison against these baselines. In the Dynamic Test-time Region Focusing module, the number of grounded region candidates is set to $M = 3$. Each candidate is augmented with zoom-in and zoom-out variants, yielding $3M + 1 = 10$ distinct visual inputs. With $P=1$ patches generated per visual input, \scheme \ produces a final candidate pool of $1 \times (3M + 1) = 10$ patches per issue, from which the final patch is selected via our automated patch selection procedure.

\noindent \textbf{Benchmark.} We evaluate \scheme\ on SWE-bench Multimodal~\cite{yang2025swebench}, the leading benchmark for assessing AI systems in fixing bugs in visual, user-facing JavaScript software. It comprises 619 task instances from 17 popular JavaScript libraries, split into a test set (SWE-bench M test, 517 instances) and a development set (SWE-bench M dev, 102 instances). We use SWE-bench M test for overall effectiveness evaluation and SWE-bench M dev for the generalizability study.

\subsection{Baselines}

We choose the baselines from the official SWE-Bench M Leaderboard~\cite{Leaderboard} with 10 issue repairing frameworks with 22 variants based on GPT~\cite{GPT4o, o3} and Claude~\cite{Claude35} backbones.


\begin{itemize}[leftmargin=*]

    \item RAG~\cite{yang2025swebench} uses BM25~\cite{robertson2009probabilistic} for document retrieval while inheriting their document formatting and prompt structure.


    \item SWEAgent~\cite{yang2024sweagent} is a lightweight framework that connects an LM to an operating system via a text-based agent-computer interface (ACI), enabling file editing and shell execution.
    

    \item Agentless~\cite{xia2025demystifying} employs a simplistic three-phase process of localization, repair, and patch validation.


    \item Agentless Lite~\cite{AgentlessLite} is a lightweight variant of Agentless that retrieves relevant files via embeddings and iteratively generates repair patches from the top results.


    \item Globant Code Fixer Agent~\cite{Globant} is an AI-driven development agent designed to automatically detect, analyze, and fix code issues by leveraging large language models to support automated software maintenance workflows.

    \item Zencoder~\cite{Zencoder} is a coding assistant that helps developers generate, edit, and understand code through natural language interactions integrated into development workflows.
    
    \item Openhands-Versa~\cite{wang2024openhands} is a generalist agent built with a modest number of general tools: code editing and execution, web search, and multimodal web browsing.

    \item Refact AI Agents~\cite{refact} are end-to-end software engineering agents that integrate with developer tools, understand codebases, and autonomously plan, execute, and iterate on complex coding tasks until completion.
    
    \item GUIRepair~\cite{huang2025seeing} applies the image2code and code2image pipeline to achieve the cross-modality reasoning. 

    \item SVRepair~\cite{tang2026svrepair} consists of a fine-tuned vision-language model that translates visual artifacts into structured symbolic representations, and an intelligent refinement loop that iteratively improves repair quality.
\end{itemize}

\subsection{Research Questions}

\begin{table*}[ht]
\centering
\caption{Results on SWE-bench M test. \raisebox{-0.15em}{\includegraphics[height=0.9em]{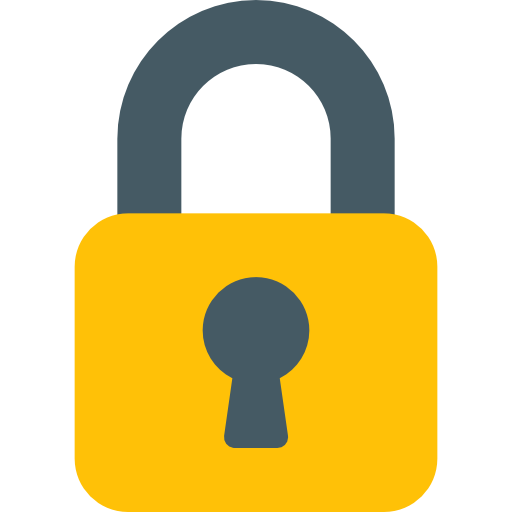}} refers closed-source systems. 
\raisebox{-0.20em}{\includegraphics[height=1em]{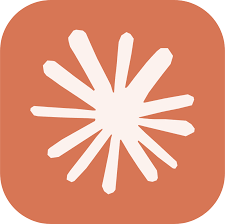}} refers claude models. 
\raisebox{-0.22em}{\includegraphics[height=1em]{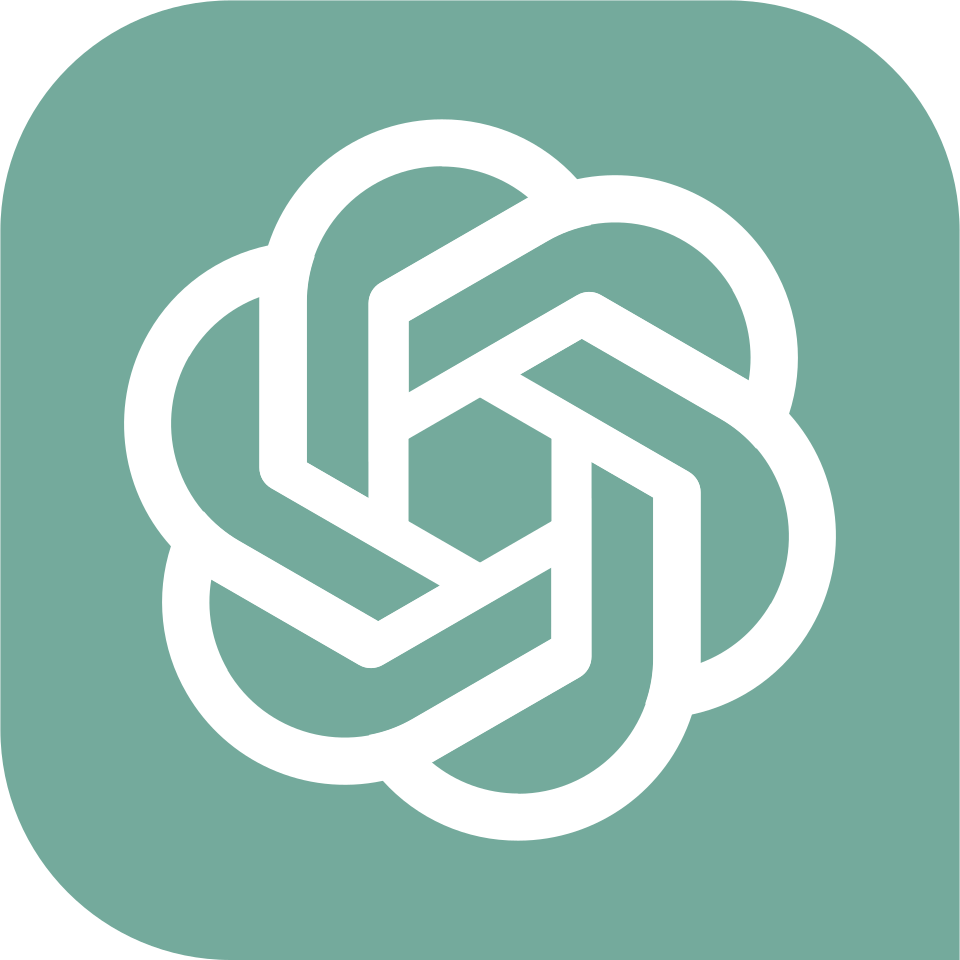}} refers gpt models.}
\label{tab:result_RQ1}
\begin{tabular}{llccc}
\toprule
\rowcolor[HTML]{C8E6C9}
\multicolumn{1}{c}{\textbf{Methods}} & \multicolumn{1}{c}{\textbf{Base Model}} & \textbf{Resolved} & \textbf{\%Resolved} & \textbf{\$Avg. Cost} \\
\midrule

\raisebox{-0.15em}{\includegraphics[height=1em]{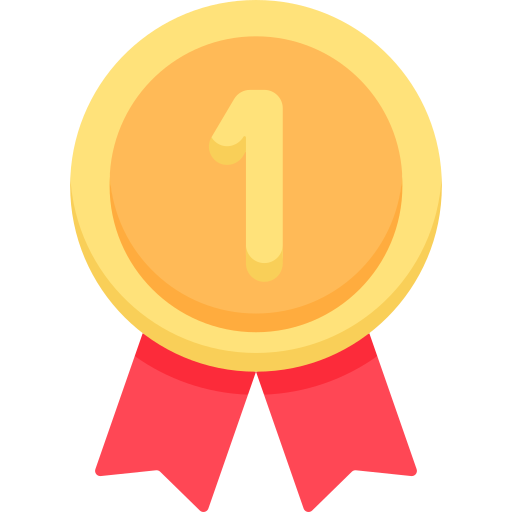}} GUIRepair~\cite{huang2025seeing}
    & {\includegraphics[height=1em]{figure/gpt.png}}~o3-20250416
    & 186 & 35.98\% & {---} \\
\rowcolor[HTML]{F1F8F1}
\raisebox{-0.15em}{\includegraphics[height=1em]{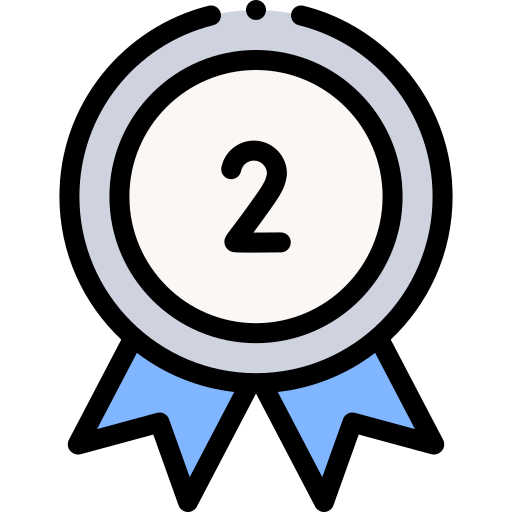}} SVRepair~\cite{tang2026svrepair}
    & {\includegraphics[height=1em]{figure/gpt.png}}~o3-20250416
    & 186 & 35.98\% & {---} \\
\raisebox{-0.15em}{\includegraphics[height=1em]{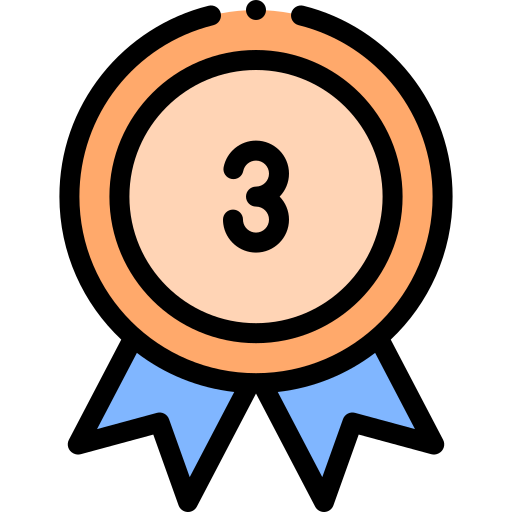}} Refact.ai Agent~\cite{refact}
    & NA
    & 184 & 35.59\% & {---} \\
\rowcolor[HTML]{F1F8F1}
OpenHands-Versa~\cite{wang2024openhands}
    & {\includegraphics[height=1em]{figure/claude.png}}~Claude-Sonnet-4-20250514
    & 178 & 34.43\% & \$1.79 \\
GUIRepair~\cite{huang2025seeing}
    & {\includegraphics[height=1em]{figure/gpt.png}}~o4-mini-20250416
    & 175 & 33.85\% & \$0.36 \\
\rowcolor[HTML]{F1F8F1}
OpenHands-Versa~\cite{wang2024openhands}
    & {\includegraphics[height=1em]{figure/claude.png}}~Claude-Sonnet-3-7-20250219
    & 162 & 31.33\% & \$1.95 \\
GUIRepair~\cite{huang2025seeing}
    & {\includegraphics[height=1em]{figure/gpt.png}}~GPT-4.1-20250414
    & 161 & 31.14\% & \$0.25 \\
\rowcolor[HTML]{F1F8F1}
Zencoder~\cite{Zencoder}~\raisebox{-0.15em}{\includegraphics[height=0.9em]{figure/lock.png}}
    & NA
    & 158 & 30.56\% & {---} \\
GUIRepair~\cite{huang2025seeing}
    & {\includegraphics[height=1em]{figure/gpt.png}}~GPT-4o-20240806
    & 157 & 30.37\% & \$0.29 \\
\rowcolor[HTML]{F1F8F1}
Globant Code Fixer Agent~\cite{Globant}~\raisebox{-0.15em}{\includegraphics[height=0.9em]{figure/lock.png}}
    & NA
    & 153 & 29.59\% & {---} \\

\midrule
Agentless Lite~\cite{AgentlessLite}
    & {\includegraphics[height=1em]{figure/gpt.png}}~o3-20250416
    & 137 & 26.50\% & - \\
    
Agentless Lite~\cite{AgentlessLite}
    & {\includegraphics[height=1em]{figure/claude.png}}~Claude-Sonnet-3-5-20240620
    & 131 & 25.34\% & \$0.25 \\
\rowcolor[HTML]{F1F8F1}
Agentless Lite~\cite{AgentlessLite}
    & {\includegraphics[height=1em]{figure/gpt.png}}~GPT-4o-20240806
    & 127 & 24.56\% & \$0.18 \\
SWE-agent Multimodal~\cite{yang2025swebench}
    & {\includegraphics[height=1em]{figure/gpt.png}}~GPT-4o-20240806
    & 63  & 12.19\% & \$2.94 \\
\rowcolor[HTML]{F1F8F1}
SWE-agent~\cite{yang2024sweagent}
    & {\includegraphics[height=1em]{figure/claude.png}}~Claude-Sonnet-3-5-20240620
    & 63  & 12.19\% & \$1.52 \\
SWE-agent JavaScript~\cite{yang2025swebench}
    & {\includegraphics[height=1em]{figure/claude.png}}~Claude-Sonnet-3-5-20240620
    & 62  & 11.99\% & \$3.11 \\
\rowcolor[HTML]{F1F8F1}
SWE-agent~\cite{yang2024sweagent}
    & {\includegraphics[height=1em]{figure/gpt.png}}~GPT-4o-20240806
    & 62  & 11.99\% & \$2.07 \\
SWE-agent Multimodal~\cite{yang2025swebench}
    & {\includegraphics[height=1em]{figure/claude.png}}~Claude-Sonnet-3-5-20240620
    & 59  & 11.41\% & \$3.11 \\
\rowcolor[HTML]{F1F8F1}
SWE-agent JavaScript~\cite{yang2025swebench}
    & {\includegraphics[height=1em]{figure/gpt.png}}~GPT-4o-20240806
    & 48  & 9.28\%  & \$0.99 \\
Agentless~\cite{xia2025demystifying}
    & {\includegraphics[height=1em]{figure/claude.png}}~Claude-Sonnet-3-5-20240620
    & 32  & 6.19\%  & \$0.42 \\
\rowcolor[HTML]{F1F8F1}
RAG~\cite{yang2025swebench}
    & {\includegraphics[height=1em]{figure/gpt.png}}~GPT-4o-20240806
    & 31  & 6.00\%  & \$0.17 \\
RAG~\cite{yang2025swebench}
    & {\includegraphics[height=1em]{figure/claude.png}}~Claude-Sonnet-3-5-20240620
    & 26  & 5.03\%  & \$0.15 \\
\rowcolor[HTML]{F1F8F1}
Agentless~\cite{xia2025demystifying}
    & {\includegraphics[height=1em]{figure/gpt.png}}~GPT-4o-20240806
    & 16  & 3.09\%  & \$0.38 \\

\midrule

\rowcolor[HTML]{C8E6C9}
\raisebox{-0.23em}{\includegraphics[height=1em]{figure/crab.png}}~\scheme~\textbf{(Ours)}
    & {\includegraphics[height=1em]{figure/gpt.png}}~o3-20250416
    & \textbf{196} & \textbf{37.91\%} & \$0.47 \\
\bottomrule
\end{tabular}
\end{table*}

\begin{table}[t]
\caption{Comparison of resolved issues across repositories between \scheme \ and  Top-4 methods on SWE-Bench M test. Backbone models are o3-20250416 for gpt and Claude-Sonnet-4-20250514 for claude. Best results are marked by green.}
\label{tab:experiment_by_repo}
\centering
\resizebox{\linewidth}{!}{
\setlength{\tabcolsep}{1pt}
\begin{tabular}{lc|c|c|c|c|c}
\toprule
\textbf{Repo} & \textbf{Num}
& {\includegraphics[height=1em]{figure/gpt.png}} \textbf{VisualRepair}
& {\includegraphics[height=1em]{figure/gpt.png}} \textbf{GUIRepair~\cite{huang2025seeing}}
& {\includegraphics[height=1em]{figure/gpt.png}} \textbf{SVRepair~\cite{tang2026svrepair}}
& \textbf{Refact~\cite{refact}}
& {\includegraphics[height=1em]{figure/claude.png}} \textbf{OpenHands-Versa~\cite{wang2024openhands}} \\
\midrule
next         & 39  & \cellcolor[HTML]{C8E6C9} 10 (25.64\%) & \cellcolor[HTML]{C8E6C9} 10 (25.64\%) & \cellcolor[HTML]{C8E6C9} 10 (25.64\%) & 8 (20.51\%)  & 9 (23.08\%)  \\
bpmn-js      & 54  & \cellcolor[HTML]{C8E6C9} 41 (75.93\%) & 37 (68.52\%) & 38 (70.37\%) & 35 (64.81\%) & 37 (68.52\%)  \\
carbon       & 133 & \cellcolor[HTML]{C8E6C9}  28 (21.05\%) & 24 (18.05\%) & 24 (18.05\%) & 26 (19.55\%) & 21 (15.79\%)  \\
eslint       & 11  & \cellcolor[HTML]{C8E6C9} 6 (54.55\%) & \cellcolor[HTML]{C8E6C9} 6 (54.55\%)  & \cellcolor[HTML]{C8E6C9} 6 (54.55\%)  & 3 (27.27\%)  & 4 (36.36\%)    \\
lighthouse   & 54  & \cellcolor[HTML]{C8E6C9} 11 (20.37\%) & 8 (14.81\%)  & 9 (16.67\%)  & 7 (12.96\%)  & 5 (9.26\%)     \\
grommet      & 20  & \cellcolor[HTML]{C8E6C9} 5 (25.00\%) & \cellcolor[HTML]{C8E6C9} 5 (25.00\%)  & 4 (20.00\%)  & 3 (15.00\%)  & 2 (10.00\%)    \\
highlight.js & 39  & 4 (10.26\%) & 3 (7.69\%)   & 4 (10.26\%)  & \cellcolor[HTML]{C8E6C9} 5 (12.82\%)  & \cellcolor[HTML]{C8E6C9} 5 (12.82\%)    \\
openlayers   & 79  & \cellcolor[HTML]{C8E6C9} 77 (97.47\%) & 76 (96.20\%) & \cellcolor[HTML]{C8E6C9} 77 (97.47\%) & 76 (96.20\%) & 74 (93.67\%)  \\
prettier     & 13  & 3 (23.08\%) & 4 (30.77\%)  & 4 (30.77\%)  & \cellcolor[HTML]{C8E6C9} 6 (46.15\%)  & 4 (30.77\%)    \\
prism        & 38  &  10 (26.32\%) & 8 (21.05\%)  & 8 (21.05\%)  & 11 (28.95\%) & \cellcolor[HTML]{C8E6C9} 12 (31.58\%)   \\
quarto-cli   & 24  & 1 (4.17\%) & \cellcolor[HTML]{C8E6C9} 5 (20.83\%)  & 1 (4.17\%)   & 4 (16.67\%)  & \cellcolor[HTML]{C8E6C9} 5 (20.83\%)   \\
scratch-gui  & 6   & 0 (0.00\%) & 0 (0.00\%)   & \cellcolor[HTML]{C8E6C9} 1 (16.67\%)  & 0 (0.00\%)   & 0 (0.00\%)      \\
\midrule
\textbf{Total}   & 517 & \cellcolor[HTML]{C8E6C9} 196 (37.91\%) & 186 (35.98\%) & 186 (35.98\%) & 184 (35.59\%) & 178 (34.43\%) \\
\bottomrule
\end{tabular}
}
\end{table}

\begin{itemize}[leftmargin=*]
    \item (RQ1) \textbf{Overall Performance.} How well does the overall performance of \scheme \ compare to SOTA baselines? 
    
    
    \item (RQ2) \textbf{Ablation Study.} How does the key components of \scheme \ contribute to the overall repair capability?

    \item (RQ3) \textbf{Parameter Study.} How do key parameters affect the performance of \scheme?
    
    
    
    \item (RQ4) \textbf{Generalizability Study}: Can \scheme \ generalize to additional multimodal task instances? 

    
    \item (RQ5) \textbf{Case Study}: Why does \scheme \ work?
    
\end{itemize}

\section{Experiment Results}

\subsection{Overall Performance (RQ1)}

We run \scheme\ three times and report the average results, with a variance of 1. Table~\ref{tab:result_RQ1} shows the performance of all approaches on the SWE-bench Multimodal test set. \scheme, using o3-20250416 as the backbone model, resolves 196 instances, achieving a resolve rate of 37.91\%, outperforming all existing approaches on performance while maintaining low cost at an average of \$0.47 per issue.

\textbf{Comparison with open-source systems.} \scheme \ outperforms all open-source baselines by a significant margin. The strongest open-source competitors, GUIRepair and SVRepair, both using o3 as backbone model, resolve 186 instances (35.98\%), which \scheme \ exceeds by 10 instances. Among other open-source systems, Agentless Lite with o3-20250416 achieves 26.50\% (137 instances), and OpenHands-Versa with Claude-Sonnet-4 reaches 34.43\% (178 instances), both of which \scheme \ surpasses by a substantial margin. Notably, \scheme \ achieves these results at an average cost of \$0.47, far lower than comparable baselines such as OpenHands-Versa (\$1.79) and SWE-agent (\$1.52--\$3.11), demonstrating a favorable balance between repair effectiveness and computational efficiency.

\textbf{Comparison with closed-source commercial systems.} Among the two closed-source commercial systems, Zencoder resolves 158 instances and Globant Code Fixer Agent resolves 153 instances. \scheme \ outperforms both by a considerable margin of 36 and 41 instances respectively, validating the effectiveness of \scheme \ across both open-source and commercial system comparisons.


\textbf{Repository-level analysis.} Table~\ref{tab:experiment_by_repo} presents the repository-level performance of \scheme \ and the top-4 baseline approaches on the SWE-bench Multimodal test set. \scheme \ achieves the best or tied-best performance across the majority of repositories. On bpmn-js~\cite{BPMN} and carbon~\cite{carbon-design-system}, \scheme \ achieves the highest resolve rate of 75.93\% and 21.05\%, respectively, where issue reports frequently contain GIF-based interaction sequences, the GIF tool in ITTC effectively extracts key frames to capture dynamic bug behaviors, contributing to this strong performance. On openlayers~\cite{openlayers} and lighthouse~\cite{Lighthouse}, where bug screenshots are particularly noisy and contain large bug-irrelevant regions, DTRF's multi-granularity region focusing proves critical, with \scheme \ resolving 77 instances (97.47\%) and 11 instances (20.37\%) respectively, outperforming all baselines on both repositories. On prism~\cite{PrismJS}, the code library tool provides matching syntax highlighting templates that facilitate accurate bug reproduction and patch validation, enabling \scheme \ to outperform GUIRepair and SVRepair with a resolve rate of 26.32\%.

\begin{figure}[h]
    \centering
    \includegraphics[width = .35\textwidth]{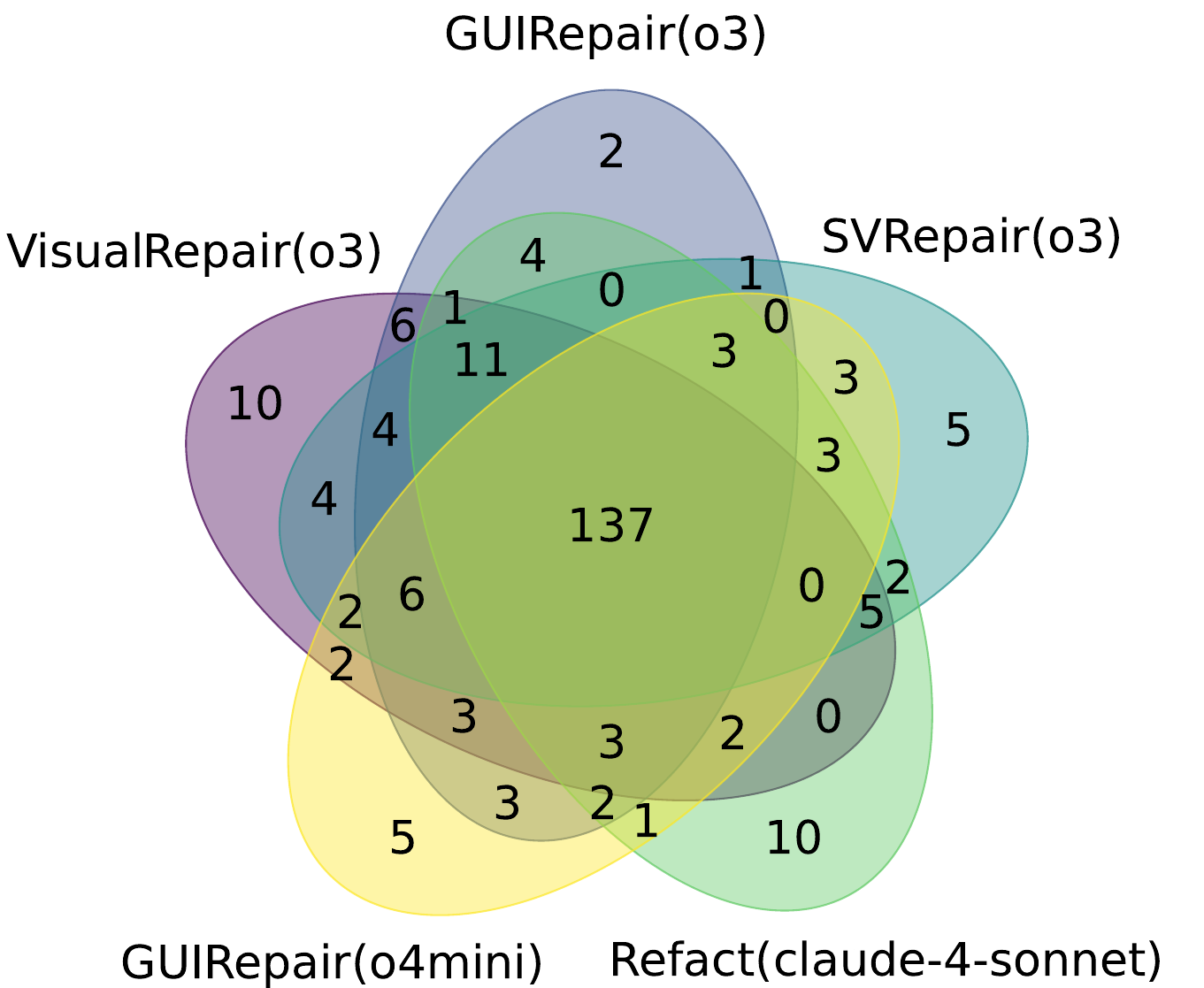}
    \caption{Venn diagram of fixed issues for Top-5 methods.}
    \label{fig:venn_top5}
\end{figure}

\textbf{Repair Difference Analysis.} Figure~\ref{fig:venn_top5} shows a Venn diagram of resolved issues across the top-5 approaches. All methods share a common core of 137 issues, representing straightforward bugs. Notably, \scheme\ uniquely resolves 10 instances, the highest among all methods, indicating its ability to capture complementary bug patterns via type-aware visual understanding and region-focused localization. In comparison, GUIRepair (o3), SVRepair (o3), GUIRepair (o4-mini), and Refact (claude-4-sonnet) uniquely resolve 2, 5, 5, and 10 instances, respectively. The limited pairwise-only overlaps between \scheme\ and others further suggest that its gains are not subsumed by existing approaches, but instead provide genuinely complementary repair capabilities.


In summary, these results confirm that addressing the heterogeneity of visual inputs and the imprecision of bug region localization leads to significant performance improvements for visual software issue repair.

\subsection{Ablation Study (RQ2)}

\scheme \ comprises two core components: Image Type-aware Tool Calling (ITTC) and Dynamic Test-time Region Focusing (DTRF). We use four variants of \scheme \ ($C_0$--$C_3$) to evaluate the individual contribution of each component, where $C_0$ serves as the baseline with both modules removed, and $C_3$ represents the full configuration.
\begin{table}[h]
\centering
\caption{The performance of 4 variants for \scheme. {\color[HTML]{FF0000} X} denotes removal and {\color[HTML]{00CD66} Y} denotes inclusion.}
\label{tab:ablation}
\begin{tabular}{@{}ccc|ccccc@{}}
\toprule
ITTC & DTRF & & Resolved & \%Resolved & \$Avg.Cost \\
\midrule
{\color[HTML]{FF0000} X} & {\color[HTML]{FF0000} X}  & $C_0$ & 151 & 29.21\% & \$0.16 \\
{\color[HTML]{00CD66} Y} & {\color[HTML]{FF0000} X}  & $C_1$ & 171 & 33.08\% & \$0.45 \\
{\color[HTML]{FF0000} X} & {\color[HTML]{00CD66} Y} & $C_2$ & 166 & 32.11\% & \$0.46  \\
{\color[HTML]{00CD66} Y} & {\color[HTML]{00CD66} Y} & $C_3$ & \textbf{196} & \textbf{37.91\%} & \$0.47 \\
\bottomrule
\end{tabular}
\end{table}


\noindent\textbf{Contribution of ITTC.} Enabling ITTC alone ($C_1$) improves performance to 171 instances, a gain of 20 over the uniform perception baseline $C_0$, demonstrating that type-aware visual interpretation substantially enhances comprehension of heterogeneous bug images. The gains are concentrated in visually heterogeneous types: UI (7), GIF (5), IDE Code (4), and Text (4), confirming that type-aware routing is most critical for inputs that a uniform strategy systematically mishandles.


\noindent\textbf{Contribution of DTRF.} Enabling DTRF alone ($C_2$) resolves 166 instances, a gain of 15 over $C_0$, validating that multi-granularity region focusing improves fault localization and patch diversity. Manual inspection confirms grounding quality: 14 of 15 newly resolved instances (93.3\%) had the bug-relevant region correctly identified, indicating that DTRF's region focusing directly drives repair rather than incidentally benefiting from patch diversity alone.



\noindent\textbf{Synergy of ITTC and DTRF.} Finally, the full \scheme \ configuration ($C_3$) achieves the best performance of 196 instances, outperforming $C_1$ and $C_2$ by 25 and 30 instances respectively, confirming that ITTC and DTRF are complementary and mutually reinforcing.

\subsection{Parameter Study (RQ3)}

\subsubsection{The $M$ (number of grounded regions) and patch generation count $P$} Rather than expensive grid search, our design was guided by prior work: GUIRepair and Agentless generate 40 patches per issue, so setting $M=3$ with 4 patches per region would yield a comparable total. However, we found that 40 patches impose a substantial selection burden on the MLLM ranker, so we reduced $P$ to 10 (approximately 1 patch per region per zoom level), balancing patch diversity with selection tractability. We sample 100 test instances for \textbf{sensitivity analysis} and found that resolved counts remain stable when $M$ ranges from 1 to 3 ($P$ from 4 to 10). As shown in Table~\ref{tab:param_m}, $M=3$ with $P=10$ achieves the highest resolved count (37), while $M=4$ ($P=13$) yields diminishing returns (dropping to 33), as additional regions increasingly overlap with already-covered areas and the larger candidate pool further burdens patch selection. Notably, $P=10$ is more efficient than GUIRepair's 40 patches, reducing inference cost and selection complexity without sacrificing performance.


\begin{table}[h]
\centering
\begin{minipage}{0.2\textwidth}
    \centering
    \caption{Sensitivity analysis of $M$ and $P$.}
    \label{tab:param_m}
    \resizebox{\textwidth}{!}{%
    \setlength{\tabcolsep}{1pt}
    \begin{tabular}{lcccc}
    \toprule
    $M$ (\#Regions) & 1 & 2 & 3 & 4 \\
    \midrule
    $P$ (\#Patches) & 4 & 7 & 10 & 13 \\
    Solved          & 35 & 36 & 37 & 33 \\
    \bottomrule
    \end{tabular}}
\end{minipage}
\begin{minipage}{0.27\textwidth}
    \centering
    \caption{Coverage and overlap ratio under different zoom factors.}
    \label{tab:param_zoom}
    \resizebox{\textwidth}{!}{%
    \setlength{\tabcolsep}{1pt}
    \begin{tabular}{lccc}
    \toprule
    Factor & Coverage\_Ratio & Overlap\_Ratio & Solved \\
    \midrule
    No\_zoom & 0.3459 & 0.1275 & 34\\
    7/8      & 0.3730 & 0.1458 & 34 \\
    1/2      & 0.4917 & 0.2384 & 37 \\
    1/8      & 0.8093 & 0.6124 & 35\\
    \bottomrule
    \end{tabular}}
\end{minipage}
\end{table}

\subsubsection{The Grounding Zoom Factor} We evaluate each setting by its coverage ratio (proportion of the image captured across regions) and overlap ratio (redundancy between regions). As shown in Table~\ref{tab:param_zoom}, a zoom factor close to 1 (e.g., 7/8) produces only minor region variations, yielding low coverage (0.373) and insufficient diversity, resulting in only 34 resolved instances. Conversely, a zoom factor close to 0 (e.g., 1/8) enlarges regions aggressively, causing the overlap ratio to spike to 0.61 and severely undermining patch diversity, dropping to 35 resolved instances. A zoom factor of 1/2 achieves the best balance, with a coverage ratio of 0.492 and a moderate overlap ratio of 0.238, yielding the highest resolve count of 37 instances. We therefore select zoom 1/2 as the optimal setting.

\begin{table}[h]
\centering
\caption{Ablation study on the threshold coefficient $k$.}
\label{tab:ablation_k}
\setlength{\tabcolsep}{1pt}
\begin{tabular}{lcccccc}
\toprule
$k$ & 0.2 & 0.5 & 0.8 & 1.0 & 1.2 & 1.5 \\
\midrule
Avg.Keyframes & 12.09 & 10.24 & 8.93 & 8.43 & 7.96 & 7.04 \\
Solved          & 37 & 36 & 36 & 38 & 38 & 32 \\
\bottomrule
\end{tabular}
\end{table}

\subsubsection{The key frame extract coefficient $k$}
$k$ controls the sensitivity of key frame selection. We sample 50 GIF-based issues to evaluate the effect of $k$, with results shown in Table~\ref{tab:ablation_k}. The number of solved issues remains stable at 36--38 when $k$ ranges from 0.2 to 1.2, but drops to 32 at $k = 1.5$, indicating that overly aggressive filtering discards visually informative frames. We therefore set $k = 1.2$ as the default, which minimizes the average number of keyframes per GIF while preserving solving performance, striking the best balance between efficiency and effectiveness.


\begin{table}[h]
\caption{Results on SWE-bench M dev. The backbone models are GPT-4o-20240806. Best results are marked by green.}
\label{tab:result_RQ3_repo}
\centering
\resizebox{\linewidth}{!}{
\setlength{\tabcolsep}{1pt}
\begin{tabular}{lc|ccccc}
\toprule
\textbf{Repo} & \textbf{Num}
& \textbf{VisualRepair}
& \textbf{GUIRepair~\cite{huang2025seeing}}
&  \textbf{SWE-agentM~\cite{yang2024sweagent}}
& \textbf{AgentlessJS~\cite{AgentlessLite}}
& \textbf{RAG~\cite{yang2025swebench}} \\
\midrule
wp-calypso & 37  & 0 (0.00\%)  & 0 (0.00\%)  & 0 (0.00\%)  & 0 (0.00\%)  & \cellcolor[HTML]{C8E6C9} 1 (2.70\%)   \\
Chart.js   & 24  & \cellcolor[HTML]{C8E6C9} 8 (33.33\%) & 6 (25.00\%) & 5 (20.83\%) & 0 (0.00\%)  & 3 (12.50\%)  \\
react-pdf  & 11  & 0 (0.00\%)  & 0 (0.00\%)  & 0 (0.00\%)  & 0 (0.00\%)  & 0 (0.00\%)   \\
marked     & 14  & \cellcolor[HTML]{C8E6C9} 8 (57.14\%) & 1 (7.14\%)  & 1 (7.14\%)  & 0 (0.00\%)  & 3 (21.43\%)  \\
p5.js      & 16  & \cellcolor[HTML]{C8E6C9} 9 (56.25\%) & 7 (43.75\%) & 4 (25.00\%) & 1 (6.25\%)  & 4 (25.00\%)  \\
\midrule
\textbf{Total} & 102 & \cellcolor[HTML]{C8E6C9} \textbf{25 (24.51\%)} & 14 (13.73\%) & 10 (9.80\%) & 1 (0.98\%) & 11 (10.78\%) \\
\bottomrule
\end{tabular}}
\end{table}


\begin{table}[h]
\caption{Results on SWE-bench M dev across different backbone models. Best results per repo are marked in green.}
\label{tab:result_RQ3_backbone}
\centering
\resizebox{\linewidth}{!}{
\setlength{\tabcolsep}{2pt}
\begin{tabular}{lc|cc|cc}
\toprule
\multirow{2}{*}{\textbf{Repo}} & \multirow{2}{*}{\textbf{Num}}
& \multicolumn{2}{c|}{\textbf{{\includegraphics[height=0.7em]{figure/gpt.png}}o4-mini-2025-04-16}}
& \multicolumn{2}{c}{\textbf{{\includegraphics[height=0.7em]{figure/gpt.png}}o3-20250416}} \\
& & \textbf{GUIRepair} & \textbf{\scheme} & \textbf{GUIRepair} & \textbf{\scheme} \\
\midrule
wp-calypso & 37 & 1 (2.70\%)   & \cellcolor[HTML]{C8E6C9} 2 (5.41\%)    & 1 (2.70\%)   & 1 (2.70\%)  \\
Chart.js   & 24 & 21 (87.50\%) & 20 (83.33\%) & \cellcolor[HTML]{C8E6C9} 23 (95.83\%) & \cellcolor[HTML]{C8E6C9} 23 (95.83\%) \\
react-pdf  & 11 & 1 (9.09\%)   & 1 (9.09\%)   & 1 (9.09\%)   & \cellcolor[HTML]{C8E6C9} 2 (18.18\%) \\
marked     & 14 & 4 (28.57\%)  & \cellcolor[HTML]{C8E6C9} 5 (35.71\%)   & 4 (28.57\%)  & \cellcolor[HTML]{C8E6C9} 5 (35.71\%) \\
p5.js      & 16 & 5 (31.25\%)  & \cellcolor[HTML]{C8E6C9} 8 (50.00\%)  & 3 (18.75\%)  & 6 (37.50\%) \\
\midrule
\textbf{Total} & 102
& 32 (31.37\%) & 36 (35.29\%)
& 32 (31.37\%) & \cellcolor[HTML]{C8E6C9} 37 (36.27\%) \\
\bottomrule
\end{tabular}
}
\end{table}

\subsection{Generalization Study (RQ4)}


In the generalizability experiment, we focus on the performance of \scheme \ on five additional repositories from the SWE-bench Multimodal dev split and base models. In Table~\ref{tab:result_RQ3_repo}, \scheme~achieves the best overall performance, resolving 25 out of 102 instances (24.51\%), outperforming other baselines. Notably, \scheme~achieves particularly strong results on Chart.js~\cite{chartjs} (33.33\%) and marked~\cite{Marked} (57.14\%), where the code library tool in ITTC provides matching code templates that significantly facilitate bug reproduction and patch validation for these rendering-oriented libraries. On p5.js~\cite{P5JS}, \scheme \ also leads with 56.25\%, outperforming all baselines. Overall, the consistent performance gains across diverse repositories validate the generalizability.

\subsection{Case Study (RQ5)}



\begin{figure}[h]
    \centering
    \includegraphics[width = .4\textwidth]{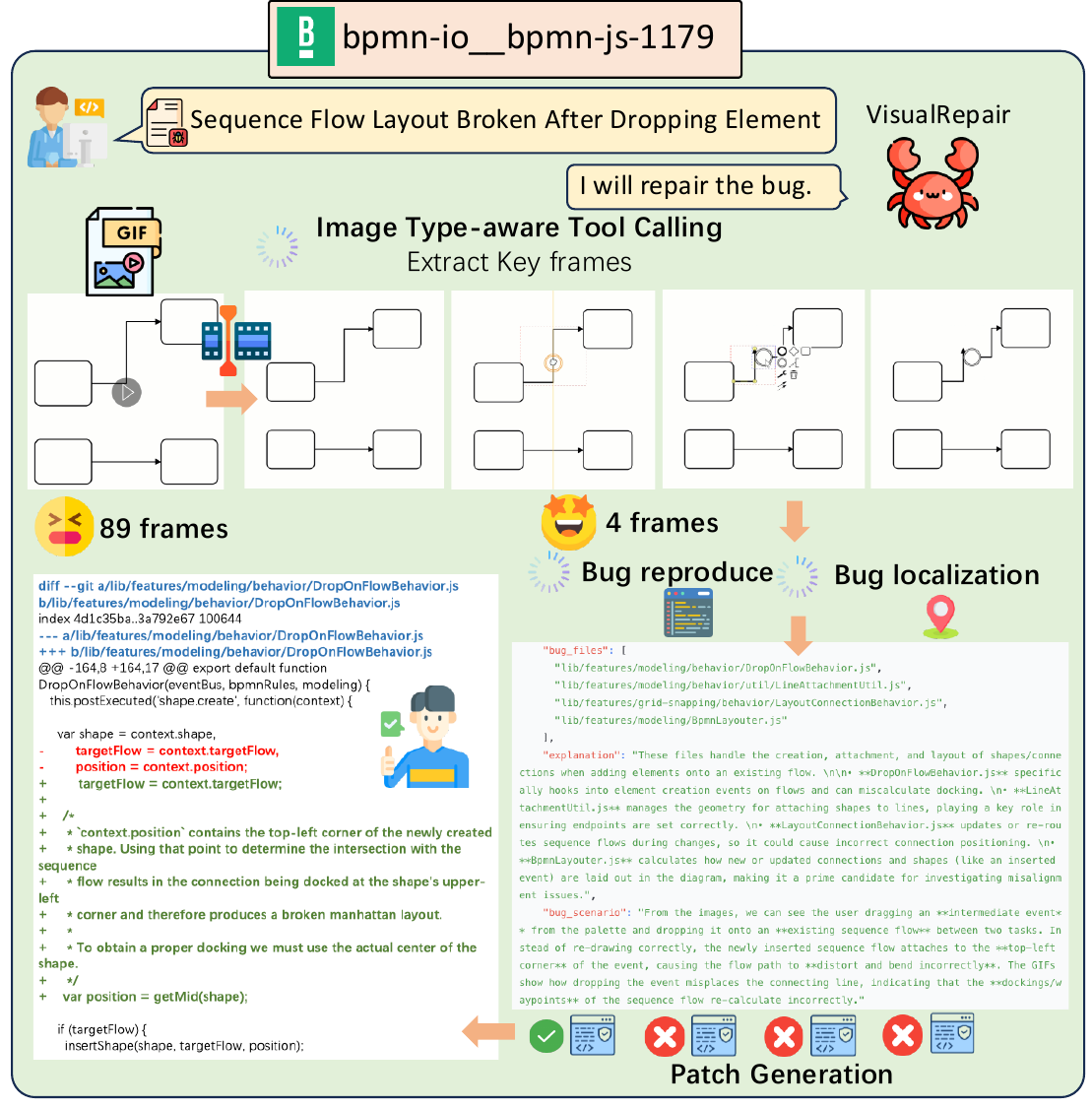}
    \caption{Case study for tool using.}
    \label{fig:tool_case}
\end{figure}

\textbf{Case Study 1: Tool using for bug comprehension.} Figure~\ref{fig:tool_case} presents a case from the bpmn-io~\cite{BPMN} repository (issue \texttt{bpmn-js-1179}), where the bug describes a sequence flow layout breaking after dropping an element, accompanied by a GIF of 89 frames. Without the GIF tool, the MLLM would struggle to extract meaningful temporal information from the raw animation. By extracting 4 representative key frames, the GIF tool reduces visual redundancy by 95\% while retaining the critical moments that reveal the erroneous behavior: the newly inserted sequence flow incorrectly attaches to the top-left corner of the dropped element rather than its center. Grounded in this compact temporal evidence, \scheme successfully localizes the bug in \texttt{DropOnFlowBehavior.js} and generates the correct patch, demonstrating the essential role of the GIF tool in comprehending dynamic bug behaviors.

\begin{figure}[h]
    \centering
    \includegraphics[width = .4\textwidth]{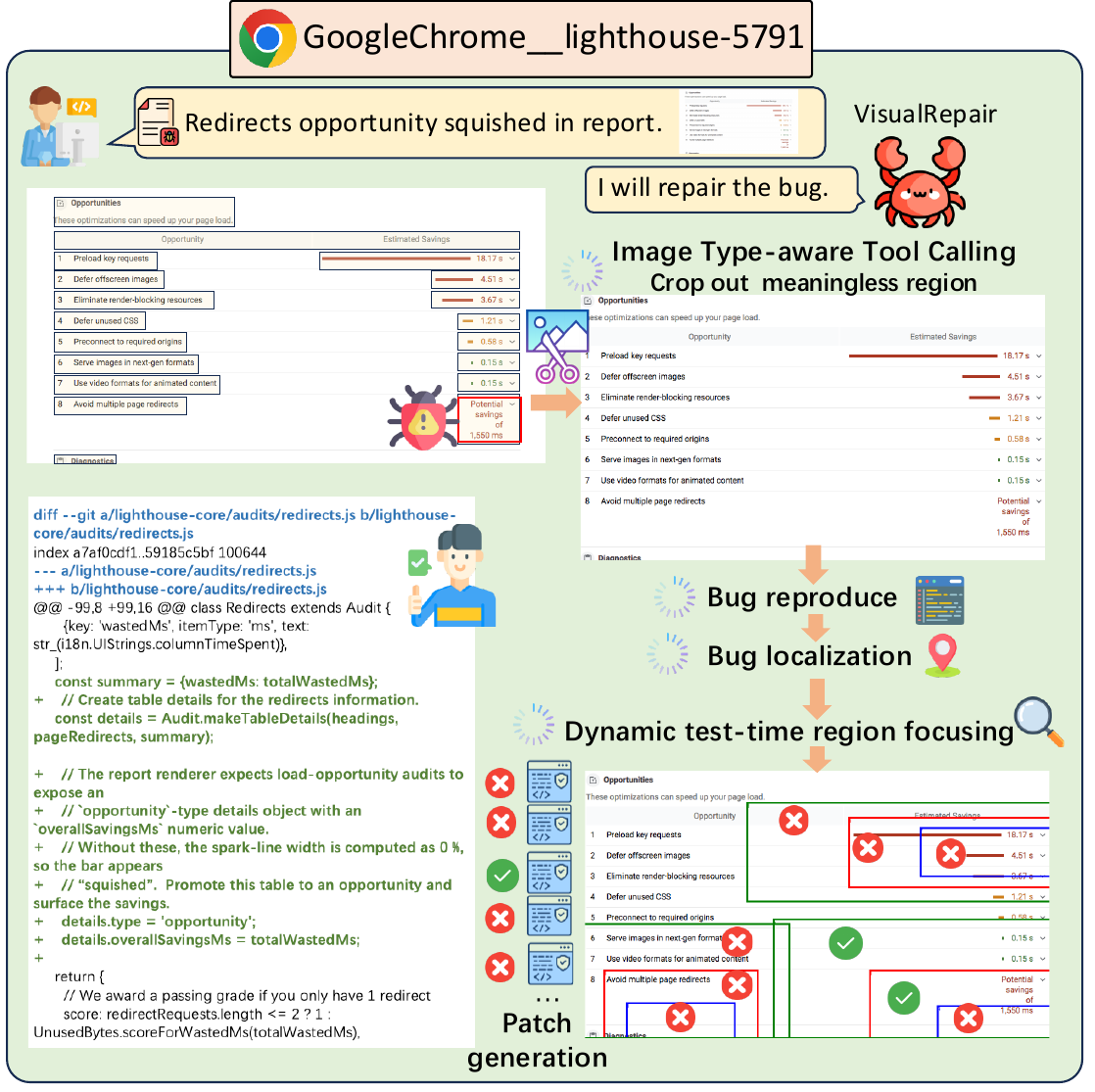}
    \caption{Case study for dynamic test-time region focusing.}
    \label{fig:grounding_case}
\end{figure}

\textbf{Case Study 2: Dynamic region grounding for localization.} 
Figure~\ref{fig:grounding_case} presents a case from the lighthouse~\cite{Lighthouse} repository (issue \texttt{lighthouse-5791}), where the bug describes redirects opportunity being incorrectly squished in the performance report. The issue image contains a cluttered audit UI with substantial uninformative whitespace, making direct bug localization difficult. \scheme first applies the crop tool to remove blank regions, sharpening the model's focus on meaningful content. Subsequently, DTRF prompts the MLLM to ground multiple region candidates, but as shown in the figure, two of the initial grounding results (red boxes) are incorrectly localized, demonstrating that single-shot grounding alone is insufficient to reliably capture the bug-relevant region at the bottom-right corner. DTRF further applies zoom-in (blue boxes) and zoom-out (green boxes) augmentations, among which the candidates marked with \raisebox{-0.23em}{\includegraphics[height=1em]{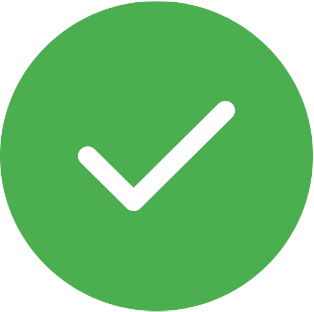}} successfully cover the defective area and provide accurate visual grounding.


\textbf{Failure Analysis}. We conduct analysis of the unresolved instances and identify three representative failure modes: (1) Long screenshots with imprecise grounding. In \texttt{carbon-design-system\_carbon-13317}~\cite{carbon-design-system}, the bug report contains an extremely long screenshot spanning multiple UI sections, making it difficult for DTRF to precisely localize the bug-relevant region. (2) GIFs with excessive frames. In \texttt{openlayers\_openlayers-9083}~\cite{openlayers}, the GIF contains an unusually large number of frames that remain numerous even after key-frame extraction, preventing the model from identifying the visual transition corresponding to the bug. (3) High-noise, high-density images. In \texttt{GoogleChrome\_lighthouse-1446}~\cite{Lighthouse}, the image contains extremely dense visual information with significant noise, preventing effective bug localization. These failure modes highlight promising directions for future work: improving visual grounding for long screenshots, developing adaptive compression methods for GIF inputs, and enhancing bug localization and patch generation under complex, noisy, or information-dense visual inputs.



\section{Threats to Validity}



(1) \textbf{Backbone Model Selection.} Our main experiments use o3 as the backbone, which may limit generalizability. However, since o3 is adopted by competitive baselines on SWE-bench Multimodal, it enables fair comparison, and additional experiments with GPT-4o and o4-mini show consistent gains. (2) \textbf{Benchmark Coverage.} We evaluate on SWE-bench Multimodal, which mainly contains JavaScript/TypeScript front-end repositories. Although performance on other languages (e.g., Java, Python, and C\#) remains unverified, it is currently the most comprehensive benchmark for visual software issue repair.

\section{Related Work}

\subsection{LLM for Automated Program Repairing}

\noindent \textbf{Single Modality Program Repair.}  Researchers have proposed numerous LLM-based APR approaches to address software engineering challenges. Early work explored two dominant paradigms: fine-tuning LLMs on bug-fix datasets~\cite{LLM4APR_Huang_ASE, LLM4APR_Jiang, FitRepair, MoRepair} and prompting pre-trained models with carefully designed templates~\cite{LLM4APR_Fan, xia2023automated, AlphaRepair, GAMMA, xia2024chatrepair, ThinkRepair, RepairAgent, DRCodePilot}, both of which demonstrated promising repair capabilities. More recently, researchers have shifted attention toward LLM-based autonomous agents capable of tackling complex, repository-level issue resolution~\cite{jimenez2024swebench, yang2024sweagent, SpecRover, xia2025demystifying, wang2024openhands}. For instance, SWE-agent~\cite{yang2024sweagent} introduces an Agent-Computer Interface to enable flexible environment interaction, SpecRover~\cite{SpecRover} enhances the repair process through automated specification inference, and Agentless~\cite{xia2025demystifying} adopts a streamlined localization-repair-validation pipeline for practical software development tasks. 

\noindent \textbf{Visual Software Issue Repair.} A separate line of work targets visual software issues; however, existing approaches are typically tailored to narrow bug categories and lack generalizability across diverse visual problem domains. Iris~\cite{Iris} proposes a context-aware technique for repairing color-related accessibility issues in mobile applications. DesignRepair~\cite{designrepair} introduces a dual-stream, knowledge-driven LLM-based framework for detecting and repairing design quality issues in front-end code. DesignBench~\cite{xiao2025designbench} benchmarks MLLMs on design-related repair across six categories of display issues. To address general real-world visual software issues, GUIRepair builds upon the agentless repair paradigm~\cite{xia2025demystifying} and incorporates cross-modal reasoning to comprehend and localize visual information. \textit{Nevertheless, existing approaches still struggle to accommodate heterogeneous visual inputs and face significant challenges in precisely grounding issue regions.}

\subsection{MLLM for GUI Development and Maintenance}

\noindent\textbf{GUI Code Generation.} Recent advances in multimodal large language models (MLLMs) have enabled increasingly sophisticated UI code generation methods. DCGen~\cite{wan2025divide} adopts a divide-and-conquer pipeline to reduce element omission and layout distortion, while LaTCoder~\cite{gui2025latcoder} improves layout fidelity through layout-aware generation. Interaction2Code~\cite{xiao2025interaction2code} evaluates MLLMs on interactive webpage generation, and EfficientUICoder~\cite{xiao2026efficientuicoder} reduces inference overhead via bidirectional token compression. Instruct4Edit~\cite{dang2025envisioning} programmatically synthesizes high-quality datasets for code editing, whereas ComUICoder~\cite{xiao2026comuicoder} uses semantic-aware segmentation and merging to enhance component reuse and maintainability.




\noindent\textbf{GUI Testing and Bug Localization.} For GUI bug localization, prior works exploit a variety of signals, including user interaction data~\cite{mahmud2024using}, natural language bug descriptions~\cite{saha2024toward}, and recorded bug videos~\cite{bernal2020translating, havranek2021v2s}, to pinpoint UI-level faults with deep learning-based methods. Building on the capabilities of LLMs, GPTDroid~\cite{liu2024make} reformulates mobile GUI testing as a question-answering task, enabling language models to interact with and explore applications through GUI information. For bug reproduction, AdbGPT~\cite{feng2024prompting} proposes a lightweight prompt engineering-based approach that automatically reproduces bugs directly from natural language bug reports. \textit{However, existing methods are either confined to code generation or applied to mobile platforms for bug reproduction, limiting their generalization to end-to-end repair of complex real-world visual software.}

\section{Conclusion}

We propose \scheme, an MLLM-based framework for visual software issue repair, addressing two key challenges: heterogeneous visual inputs and bug-region localization. \scheme\ introduces two core modules: Image Type-aware Tool Calling, which classifies input images and invokes a tailored tool chain for robust interpretation of diverse visual content, and Dynamic Test-time Region Focusing, which grounds bug-related region candidates and refines them via adaptive zoom-in/out to improve fault localization and patch diversity. Extensive experiments on SWE-bench MM demonstrate that \scheme\ consistently outperforms all leaderboard methods.






\bibliographystyle{IEEEtran} 
\bibliography{ref}

\end{document}